\documentclass[12pt]{article}
\usepackage{times}
\usepackage{geometry}
\usepackage{natbib, graphicx, floatrow}
\usepackage[utf8]{inputenc}
\usepackage{enumitem,amssymb}
\usepackage{ragged2e}
\newlist{thematic}{itemize}{8}
\setlist[thematic]{label=$\square$}
\usepackage{pifont}

\usepackage{aas_macros}

\begin{document}
\raggedright
\setlength\parindent{24pt}

\huge
Astro2020 Science White Paper \linebreak

Mapping Gas Phase Abundances and Enrichment Patterns Across Galaxy Disks \linebreak
\normalsize

\noindent \textbf{Thematic Areas:} 

$\square$ Resolved Stellar Populations and their Environments \hspace*{40pt}   

$\square$    Galaxy Evolution   \hspace*{65pt} \linebreak
  
\noindent \textbf{Principal Author: }

\noindent Name:	Kathryn Kreckel
 \linebreak						
Institution:  Max Planck Institute for Astronomy
 \linebreak
Email: kreckel@mpia.de 
 \linebreak
Phone:  +49 6221 528 451
 \linebreak
 
\noindent \textbf{Co-authors:}  
Danielle Berg (OSU), 
Guillermo A. Blanc (Carnegie Observatories), 
I-Ting Ho (MPIA), 
Bethan James (STScI),
Antonios Katsianis (U. Chile),
Juna A. Kollmeier (Carnegie), 
Mark Krumholz (ANU), 
David R. Law (STScI),
Laurie Rousseau-Nepton (CFHT),
Karin Sandstrom (UCSD),
Evan Skillman (U. Minnesota), 
Yuan-Sen Ting (IAS/Princeton/OCIW), 
Renbin Yan (U. Kentucky),
Dennis Zaritsky (U. Arizona)
  \linebreak

\noindent \textbf{Abstract:}
The distribution of gas-phase abundances in galaxy disks encodes the history of nucleosynthesis and transport through the interstellar medium (ISM) over cosmic time.  Multi-object and high resolution integral-field spectroscopy have started to measure these distributions across hundreds of HII regions individually resolved at $\lesssim 100$ pc scales in a handful of objects, but in the coming decade these studies will expand to larger samples of galaxies. This will allow us to understand the role of feedback and turbulence in driving the mixing and diffusion of metals in the ISM, and statistically assess the role of galaxy environment and disk dynamics in modifying how mixing proceeds. Detailed searches for over- and under-enriched regions can address to what extent star formation is triggered by previous generations of star formation and by pristine and recycled gas flows. Local galaxies, for which these detailed measurements will be possible within the next decade, will inform the interpretation of integrated measurements at high-z, where very different dynamical gas-rich environments are found in early disk galaxies. Currently, progress in the field is severely hampered by the 0.2-0.3 dex level systematic uncertainties plaguing nebular abundance diagnostics. Improving our detailed understanding of ionized nebulae at $<$20 pc scales will help us find a solution to this problem, which will prove key to the study of metal enrichment and mixing across the galaxy population in the next decade.


\pagebreak

\section{Important Scientific Topics in ISM Enrichment}
\vspace{-0.3cm}
The production and build up of heavy elements from stellar nucleosynthesis over cosmic time have resulted in pronounced radial trends in chemical abundances.  In the gas phase, this is generally traced through Oxygen and Nitrogen abundances.  These predominantly show decreasing abundances with increasing radius \citep{Zaritsky1994, Sanchez2014, Kaplan2016, Belfiore2017}, tracing the inside-out growth of galaxy disks \citep{Boissier1999}. 
Gas phase abundances are typically measured in HII regions \citep{Kennicutt1996}, where the young hot stars ionize their surroundings, producing ionized emission lines in multiple atomic/ionic species. 
The gas phase abundances are then inferred using methods that either directly measure the electron temperature and density of the gas (the {\it direct method}),  measure metal recombination lines (RL), or use calibrations of strong-emission-line (SEL) diagnostics. Given the relatively short lifetime of these most massive stars ($<$10 Myr), HII regions trace the instantaneous chemical abundances in the ISM. 

To date, most studies based on multi-object spectroscopy have been limited to analyzing simple radial trends due to limited statistics, with  only tens of detections per galaxy \citep{Pilyugin2014} that selectively sample only the brightest HII regions. Additionally, systematic offsets between different abundance diagnostics have plagued the interpretation and cross comparison of results \citep{Kewley2008, Blanc2015}. However, typical Milky Way mass galaxies host thousands of HII regions, and while these are often preferentially concentrated into spiral arms, they can be found throughout the galaxy disk. Wide-field high spatial resolution integral field spectroscopy allows for the efficient detection and spectral characterization of nebulae, and it is starting to systematically produce 2-dimensional maps of chemical abundances across nearby galaxies \citep{Belfiore2016, Kreckel2016, Ho2017}. 

\begin{figure}
\floatbox[{\capbeside\thisfloatsetup{capbesideposition={right,top},capbesidewidth=3.5in}}]{figure}[\FBwidth]
{\caption{A high resolution cosmological zoom simulation of a Milky Way-sized halo.  Over-density contours of the mass distribution trace the spiral pattern. The face-on colorscale map of the azimuthal residual in the metallicity field (where the radial trend has been subtracted) shows an over-density of metal rich (poor) star particles on the trailing (leading) side of the spiral arm, driven by spiral arm large-scale systematic streaming motions. Adapted from \cite{Grand2016}.}\label{fig:Grand2016}}
{\includegraphics[width=2.5in]{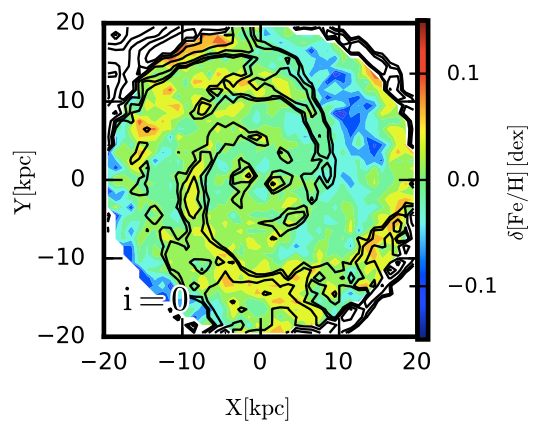}}
\end{figure}
\begin{figure}
\floatbox[{\capbeside\thisfloatsetup{capbesideposition={right,top},capbesidewidth=3in}}]{figure}[\FBwidth]
{\caption{
The Pearson correlation for metals, as a function of separation (x), produced by models of supernova-like injection events, and evaluated for parameters appropriate to local spiral galaxies, local dwarf galaxies, and z $\sim$ 2 star-forming disks. These theoretical models, based on stochastically-forced diffusion, predict the multi-scale statistics in patterns of chemical enrichment. For spiral and high-z galaxies, the correlation does not drop below 0.5 until scales larger than 500pc. 
Adapted from \cite{Krumholz2018}.}\label{fig:Krumholz2018}}
{\includegraphics[width=3in]{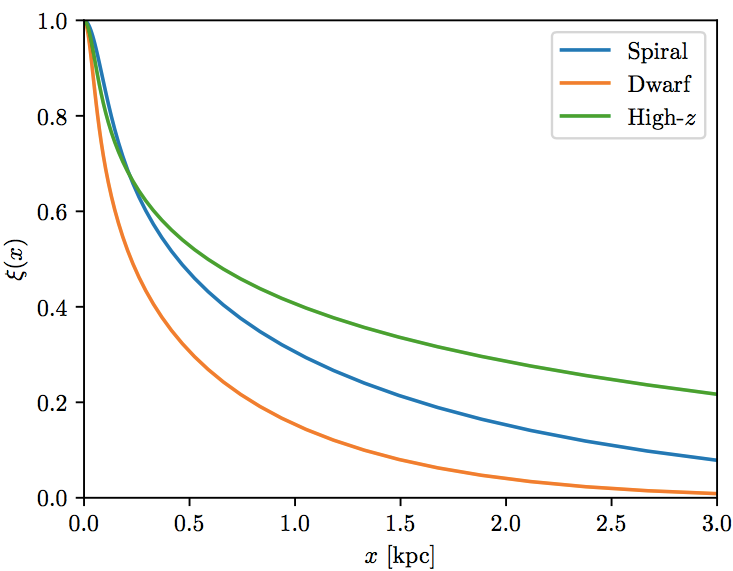}}
\end{figure}

\begin{center}
\textbf{Quantifying the mixing and diffusion in the ISM}
\end{center}
\vspace{-0.4cm}
An instantaneous picture of gas-phase abundances traces, in a statistical sense, the progression of mixing and diffusion that act to homogenize the ISM. However, that chemical transport and mixing is essentially unconstrained by existing observations on scales between individual wind-blown stellar shells ($\sim$ 1 pc; \citealt{Miller1993})
and galactic metallicity gradients ($\sim$ 1 kpc).  
This mixing is theorized to happen through a combination of bar driven radial mixing \citep{DiMatteo2013}, spiral-arm driven large-scale systematic streaming motions \citep{Grand2016, Sanchez-Menguiano2016}, kiloparsec-scale mixing-induced dilution due to the spiral density waves passage \citep{Ho2017}, thermal and gravitational instabilities \citep{Yang2012, Petit2015}, and interstellar turbulence \citep{deAvillez2002, Krumholz2018}. Each of these theories has specific predictions for the expected morphological distribution (e.g. Figure \ref{fig:Grand2016}) and spatial scales (e.g. Figure \ref{fig:Krumholz2018})  of chemical inhomogeneities within the disk, with different dependencies on local and global conditions (e.g., bar strength, star formation rate; \citealt{Roy1995}).

\begin{figure}
    \centering
    \includegraphics[height=3in]{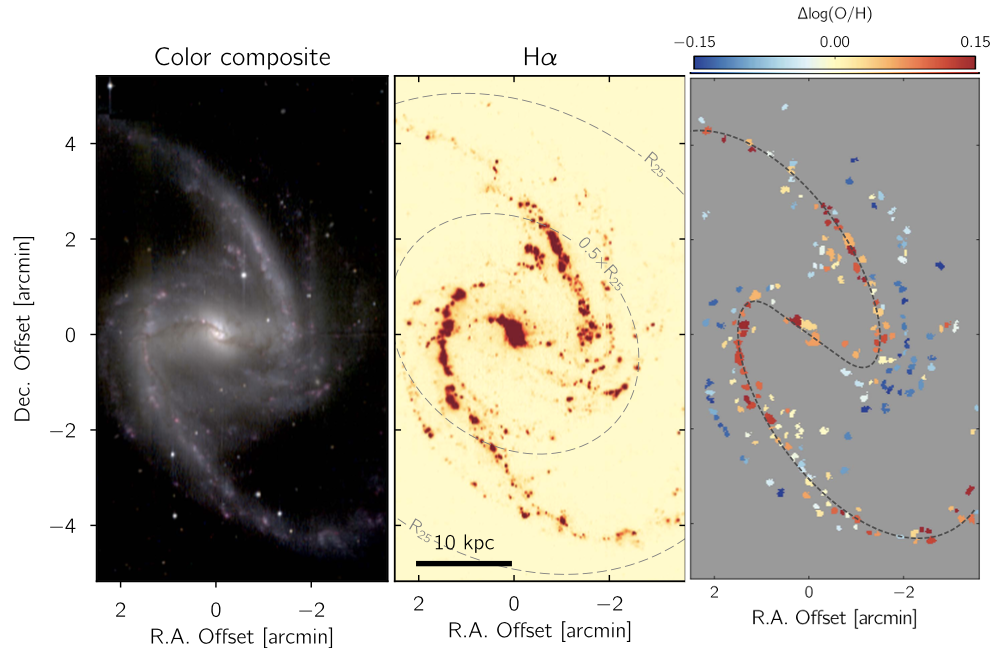}
    \caption{Composite optical image (left), H$\alpha$ emission (center) and Oxygen abundance map (right) in NGC 1365.  Here, we show the Oxygen abundance (12+log(O/H)) after subtracting the radial gradient ($\Delta$log(O/H)). Enhancement along the ridge of the spiral arm relative to both the upstream and downstream sides is apparent. Adapted from \cite{Ho2017}.}
    \label{fig:Ho2017}
\end{figure}

Observational constraints on these theories have been limited to individual objects.  Some of the earliest results reveal only marginal evidence for moderate deviations from chemical abundance homogeneity, and no evidence for significant large-scale azimuthal variations across entire galaxy disks and spiral patterns \citep{Li2013, Kreckel2016}. 
Other galaxies show more pronounced trends (Figure \ref{fig:Ho2017}), typically revealing enrichment along spiral structures  \citep{Croxall2016, Sanchez-Menguiano2016, Vogt2017, Ho2017, Ho2018}. Consistent across all observations, azimuthal variations in the abundances are found at the level of 0.1-0.2 dex, larger than what is typically predicted by simulations \citep{Roy1995, Kubryk2013, Krumholz2018}.  Optical IFU observations across the full star-forming disk on expanded galaxy samples, systematically exploring the parameter space in global galaxy properties (e.g., stellar mass, star formation rate, Hubble type), are needed to distinguish the role of these different mechanisms.  

As heavy elements play an important role in the cooling of the ISM and trace changes in fundamental local physical conditions (e.g., CO to H$_2$ conversion factor, gas to dust ratio), azimuthal variations in the chemical abundance have the potential to impact different physical processes and inferred parameters.  Trends with galaxy environment could potentially have implications on cloud scales, impacting the local star formation efficiency as as seen in galaxy-averaged trends \citep{Utomo2018}. Simulations predict that star formation is suppressed within low metallicity gas that experiences decaying turbulent conditions \citep{Walch2011}. As such, quantifying the level of abundance variations is essential for informing and testing sub-grid star formation prescriptions in simulations \citep{Shen2010, Emerick2019}.  In addition, gas phase chemical inhomogeneities could significantly impact stellar chemical tagging studies that aim to identify populations of stars that formed together in clusters.

Compared to their local descendants, galaxies at redshifts $z>0$ have systematically higher velocity dispersions, experience higher rates of low-metallicity inflow and significantly stronger outflows. These difference should all be reflected in the statistics of their abundances (Figure \ref{fig:Krumholz2018}). 
With present telescopes it has not been practical to obtain two dimensional abundance maps beyond $z \sim 2$ except in a small number of cases \citep[e.g.,][]{Law2012,Leethochawalit2016}, but this will improve dramatically with the advent of JWST and ground-based ELTs.

\begin{center}
\textbf{Localized abundance variations}  
\end{center}
\vspace{-0.4cm}Once we have constrained the expected systematic variations in abundances due to large scale dynamical processes in the disk, it is possible to search for localized deviations from these patterns.   Both extremes, towards high or low abundances, as well as large deviations from typical radial gradient slopes, inform us about processes relating to star formation, gas flows within galaxies, and gas flows into and out of the circumgalactic medium.  Searching for these outliers requires full optical IFU spectroscopic maps of galaxy disks sampling thousands of HII regions

It is an open question to what extent supernova explosions are acting to trigger or to suppress locally a future generation of star formation.  As they contribute not only  mechanical energy but also heavy elements into the surrounding gas, this localized enrichment should be visible within specific galactic dynamical environments.  For example, \cite{Ho2017} describe a simple mechanism for the observed azimuthal abundance variations in NGC 1365.  Here, simple enrichment models producing a slow localized enrichment within a single parcel of gas as it passes through the interarm environment nicely describe the `carousel'-like rise and fall of azimuthal abundances.  Specific test cases in select dynamical environments (e.g., co-rotation) will provide key insights into the case for triggered star formation. 

At the opposite extreme, regions significantly less enriched compared to their surroundings also provide interesting constraints on how star formation could be fueled.  Such anomalously low-metallicity gas requires an injection of pristine material, presumably from the circumgalactic medium, but very few studies have identified such regions \citep{Hernandez2013, James2013b, Hwang2019}.  
The presence of wide spread systematic flows of gas stimulating star formation would have far-reaching implications for our understanding of disk galaxy evolution.

\begin{center}
\textbf{Establishing precision metallicity measurements} 
\end{center}
\vspace{-0.4cm}Studies of gas phase abundances derived from strong line methods have been plagued by large uncertainties in the prescriptions used, which can vary by as much as 0.7 dex \citep{Kewley2008}.  At a fundamental level, this means we cannot definitely say if most galaxies in the local universe are sub- or super-solar metallicity.  To robustly determine the abundance within a given HII region requires knowledge of the electron density, electron temperature, and population of different ionic states. In some cases these can be directly measured, particularly where weak auroral emission lines (e.g. [OIII] $\lambda$4364, [SII] $\lambda$6312, [NII] $\lambda$5755) are observed. Using these lines, a \textit{direct method} of determining the abundances can be employed \citep{Kennicutt2003, Li2013, Berg2015}. However, these lines are faint, and fainter still in lower temperature, high metallicity HII regions.  With only the $\sim$100 times brighter strong emission lines  ([OII] $\lambda$3727, H$\beta$ $\lambda$4862, [OIII] $\lambda$5007, H$\alpha$ $\lambda$6563, [NII] $\lambda$6583, [SII] $\lambda$6716, $\lambda$6731), one must rely on \textit{strong line methods} derived from either empirical calibrations against the direct method (e.g., \citealt{Pilyugin2016}) or photo-ionization modeling (e.g., \citealt{Dopita2016}).  

The first step towards making progress on these discrepancies requires uniform observations of HII regions, where the full suite of strong and auroral lines can be measured (ideally [OII]$\lambda$3727 to [SIII]$\lambda$9069) with moderate spectral resolution (i.e. R between 2000 and 6000).  This could be enabled by deep IFU observations, using either larger telescopes (ELTs) to gain sensitivity and resolve more distant sources, or larger field of view IFUs on existing 8m class telescopes to do this more efficiently across nearby galaxies. 
These observations would allow us to standardize the model assumptions employed in the direct method and the suite of emission lines used, isolating residual dependencies on physical conditions such as the ionization parameter. 
This must be done with $\sim$100pc spatial resolution to isolate individual HII regions from the surrounding diffuse ionized gas, and for HII regions across a wide range of metallicities, which requires a galaxy sample spanning a wide range in stellar mass. Carefully constraining the fundamental biases inherent to each strong line calibration method is essential to informing our physical interpretation in the abundance variations. 
Further, resolving the internal ($<$20pc scale)  ionization and temperature structures of HII regions is needed to inform our assumptions on how these variations impact our unresolved views of most extragalactic HII regions \citep{James2013, Kumari2018}.  These scales are only available on current telescopes in the most nearby galaxies (1$^{\prime\prime}$=20pc at 4 Mpc), or in the future on an ELT with AO-fed IFU with R$\sim$4000.



\vspace{-0.3cm}
\section{Opportunities and Challenges}
\vspace{-0.3cm}
With existing seeing limited instruments ($\sim$1$^{\prime\prime}$), it is possible to achieve the physical resolutions necessary ($<$100pc) to isolate individual HII regions out to distances of $\sim$20 Mpc.  This volume encompasses thousands of galaxies.  Multiple projects are planned and underway (outlined below) that begin to address the science topics described above. However, sample sizes remain small ($<$50 galaxies) compared to the $\sim$kpc resolution of spectroscopic galaxy surveys like CALIFA \citep{Sanchez2012}, MaNGA \citep{Bundy2015} and SAMI \citep{Bryant2015}, which have hundreds to thousands of targets.  These surveys have made great progress in understanding radial gradients \citep{Sanchez2014}, but without the spatial resolution to isolate individual HII regions, detailed abundance mapping is impossible \citep{Poetrodjojo2018}.  
\vspace{-0.3cm}
\begin{itemize}[leftmargin=*]
\item TYPHOON - Longslit ``stepped-slit” or ``step-and-stare” observations of a dozen nearby galaxies. With 600s exposures on a 2.4m telescope, this survey is relatively shallow.
\vspace{-0.3cm}
\item CHAOS - Direct method abundances of $\sim$20 galaxies using LBT multi-object spectroscopy, but limited to 50-100 HII regions per galaxy.
\vspace{-0.3cm}
\item SIGNALS - SITELLE imaging Fourier transform spectroscopy of $\sim$40 galaxies.  SITELLE has a field of 11$^{\prime}$ but only strong lines within select narrow spectral windows are observed.
\vspace{-0.3cm}
\item PHANGS - VLT/MUSE IFU observations of 20 galaxies, using multiple pointings to mosaic the central star-forming disks. While MUSE has a large field of view (1$^{\prime}$), it lacks blue wavelength coverage (4800-9300\AA). Some have added coverage from KCWI  (3500-5600\AA).
\vspace{-0.3cm}
\item SDSS-V/LVM - 
IFU survey of the MW (at $<$ 1pc), Local Group (at 10-20 pc), and Local Volume (mostly D $<$ 2-4 Mpc dwarfs), with operations scheduled for 2020-2025.  This is one of the only projects achieving such high physical resolutions with full optical spectroscopy, but with 6$^{\prime\prime}$-40$^{\prime\prime}$ fibers it will predominantly target only the nearest galaxies. 
\end{itemize}
\vspace{-0.25cm}
These projects all make important progress in different ways, pushing deep enough to survey a set of galaxies and get the auroral lines (CHAOS, PHANGS, LVM), pushing wide enough to sample thousands of HII regions (TYPHOON, PHANGS, SIGNALS, LVM), and pushing to high enough spatial resolution to resolve the internal structure of star forming regions (SIGNALS, LVM).  The most recent advances in this field have come from the development of new wide-field optical integral field unit (IFU) spectrographs.  While longslit and multi-object spectrographs are available on most telescopes, any campaign to survey a large area within nearby galaxies requires a substantial investment of telescope time.  Optical IFUs are more efficient, however the leading wide-field instruments (MUSE on the VLT, KCWI on Keck) have no or limited open time available to the US community.  New optical IFUs currently under development (LVM for APO/LCO, LLAMAS for Magellan, KCRM for Keck) will not improve this situation.

\vspace{-0.35cm}
\section{Recommendations}
\vspace{-0.3cm}
New and upcoming wide-field, high resolution, sensitive optical IFU instrumentation will/would place abundance mapping in the realm of possibility for a large sample of disk galaxies.  A sufficiently wide ($>$1 arcmin$^2$) field of view enables efficient surveying of full galaxy disks, obtaining statistics for thousands of HII regions per galaxy. Such uniform, large samples are key to informing developments and improvements in photoionization modeling, which will, in turn, serve to better inform our measurements of the gas phase abundances. IFU coverage also removes the aperture biases introduced by fixed width slits, and selection biases introduced from small HII region samples.  Full optical spectroscopy, in addition to providing crucial emission lines from [OII]$\lambda$3727 to [SIII]$\lambda$9069, provides a wealth of additional information on the underlying stellar population, the stellar and gas kinematics, the ionizing and feedback sources (e.g., supernova remnants and massive Wolf Rayet stars with strong winds), and diffuse ionized gas. \looseness=-2

Crucial to these studies is our ability to resolve individual HII regions from the surrounding diffuse ISM.  At 1$^{\prime\prime}$ seeing, we can achieve this goal (roughly $<$100pc spatial scales) out to distances of 20 Mpc, and typical AO improvments double this distance.  More challenging is to resolve the internal structure, such as LVM aims to do within the Local Group, but at 1$^{\prime\prime}$ this limits us to those galaxies within 4 Mpc.  
An ELT with AO-fed IFU - for example - or a wide-field IFU on an ELT would be particularly helpful for some of the science cases detailed here. This would enable detection of faint auroral lines within metal rich HII regions, high ($<$20pc) spatial resolution  observations out to significantly larger distances, and even resolutions $\sim 100-200$ pc in the high-redshift universe.


\pagebreak

\bibliographystyle{apj}
\bibliography{refs}

\end{document}